\documentclass[12pt,manuscript]{aastex}
\usepackage{graphicx}
\usepackage{epsfig}

\begin{document}

\title{Automatic Detection of Expanding HI Shells Using Artificial Neural Networks}

\author{Anik Daigle and Gilles Joncas}
\affil{D\'{e}partement de Physique, Universit\'{e} Laval,
Qu\'{e}bec, Qc G1K 7P4 Canada, adaigle@phy.ulaval.ca,
joncas@phy.ulaval.ca}

\author{Marc Parizeau}
\affil{D\'{e}partement de G\'{e}nie \'{e}lectrique et de G\'{e}nie
Informatique, Universit\'{e} Laval, Qu\'{e}bec, Qc G1K 7P4 Canada,
parizeau@phy.ulaval.ca}

\and

\author{Marc-Antoine Miville-Desch\^enes}
\affil{Canadian Institute for Theoretical Astrophysics, 60
St-George street, Toronto, Ontario M5S 3H8 Canada,
mamd@cita.utoronto.ca}

\begin{abstract}
The identification of expanding HI shells is difficult because of
their variable morphological characteristics. The detection of HI
bubbles on a global scale therefore never has been attempted. In
this paper, an automatic detector for expanding HI shells is
presented. The detection is based on the more stable
\emph{dynamical} characteristics of expanding shells and is
performed in two stages. The first one is the recognition of the
dynamical signature of an expanding bubble in the velocity
spectra, based on the classification of an artificial neural
network. The pixels associated with these recognized spectra are
identified on each velocity channel. The second stage consists in
looking for concentrations of those pixels that were firstly
pointed out, and to decide if they are potential detections by
morphological and 21-cm emission variation considerations. Two
test bubbles are correctly detected and a potentially new case of
shell that is visually very convincing is discovered. About
$0.6\%$ of the surveyed pixels are identified as part of a bubble.
These may be false detections, but still constitute regions of
space with high probability of finding an expanding shell. The
subsequent search field is thus significantly reduced. We intend
to conduct in the near future a large scale HI shells detection
over the Perseus Arm using our detector.
\end{abstract}

\keywords{ISM: bubbles -- radio lines: ISM -- techniques: image
processing}

\section{Introduction}

The study of neutral hydrogen bubbles is intimately related to our
knowledge of the interstellar medium (ISM) and its different
phases. The evolution of bubbles is indeed function of parameters
such as the filling factors of the cold and hot phases, the number
and distribution of supernovae (SN) and Wolf-Rayet (WR) stars in
the Galaxy, etc. In order to develop a significant knowledge of
the parameters related to the physics of bubble expansion, one
needs to analyze many observational cases. There are very few
known expanding HI structures; small bubbles associated to
individual stars are particularly unheard of because of the
insufficient resolution of the HI surveys previous to the Canadian
Galactic Plane Survey (CGPS, see section 2). However, the CGPS
data have sufficient resolution for the study of those smaller
objects. How common are those objects in reality? If they reveal
themselves as rare, could we have to conclude, for example, to the
paucity of WR stars in the Galaxy? In a more general context, we
believe that it would be interesting to trace and study small
bubbles because they are isolated, and, therefore, involve fewer
energy injection processes. The study of such cases could thus
lead to more secure conclusions and better constraints, than the
study of bigger scale structures such as superbubbles, which are
caused by multiple contributions and several stars over several
hypothetical generations.

Since the morphological characteristics of HI bubbles are very
different from one bubble to another, their detection on a
morphological basis is difficult. On the other hand, their
dynamical characteristics are much more stable: the expansion
velocity of bubbles associated to individual stars (and of radius
of a few parsecs) are generally $\sim$ 10 km s$^{-1}$
\citep{Ger99,Cap00}. Hence, the two opposite sides of the shell
have a typical maximum relative velocity of 20 km s$^{-1}$. In the
CGPS HI cubes, this results, due to the Doppler shift, in a
characteristic signature detectable in the \emph{velocity spectra}
(see Figure 1): two distinct peaks can be seen, one blueshifted
and one redshifted, and separated by approximately 20 km s$^{-1}$.
Our strategy is as follows: we seek first to detect expanding
bubbles using their dynamical characteristics, hence by detecting
the characteristic two-peaked profile in the velocity spectra
associated with bubbles. Because this is a pixel-by-pixel
approach, the bubble morphology is no more of prime importance,
and we want to use it only afterwards, as a further confirmation
to the detection. The presence of two peaks separated by 20 km
s$^{-1}$ in a velocity spectrum is indeed not sufficiently
restrictive to make a reliable bubble detection: much structural
noise, for example velocity crowding on a line of sight or the
presence of converging clouds, is likely to cause false shell
detections. As will be seen, the morphology and other indicators
are therefore used for confirmation. Also, we intend to first
confine our survey to the Perseus Arm ($l = 100^\circ$ to
$140^\circ$) in order to avoid the confusion in the
velocity-distance relation existing for $l < 90^\circ$ and $l >
270^\circ$. This way, we believe that a detector pointing to
regions of space where dynamical characteristics of expanding
shells are found performs a valuable cleaning of the data and
significantly reduces the subsequent search field.

\subsection{Expanding Structures and the Models of the ISM}

The so-called bubbles, shells or cavities, referring to the ISM
gas, are technically local lacks of gas with a boundary that is
neutral or ionized and of variable density. Because of their
expansion and of their often circular morphology, it is generally
acknowledged that those objects result from a local momentum
injection in the ISM. One of two important questions that can be
asked refers to the nature of this energy source. The second
question inquires about the effect those objects have on the
structure of the ISM. Answering these questions is related more or
less directly with the particular characteristics of the shells,
which are deduced from observational data in a variety of spectral
domains. Those characteristics are mainly: the mass of the shell,
its expansion velocity, its dimensions, the density inside and
outside of the bubble. Deriving those quantities from
observational data allows one to confront them with hydrodynamical
models of expanding bubbles in a given medium, which can be
elaborated in relation to any plausible energy injection
hypothesis and models of the ISM.

Neutral hydrogen is a major component of the galactic ISM
(M$_{HI}$ $\approx$ $2\times10^9$ M$_\odot$). \citet{Fie69}'s
pioneering work opened the way to the multi-phase models of the
ISM. However, their initial model did not take into account
violent events such as SN explosions. Those were considered by
\citet{McK77}, who introduced the hot and ionized phase of the ISM
(HIM for Hot Ionized Medium). This 3-phase model efficiently
predicts the pressure, electron density and velocity dispersion of
the ISM clouds, as well as the intensity of the soft X-rays
emission of the ISM. The values of the filling factors are however
less certain. McKee \citep{McK95} and Cox \citep{Cox95,Cox74} do
not agree about the place taken by the HIM in the ISM : McKee
believes the HIM makes up the intercloud medium, while Cox
believes it has a small filling factor and is confined to
dispersed bubbles. It is the determination of the HIM filling
factor that will settle between the different ISM models -
bringing the important link between the study of expanding shells
and large scale modeling of the ISM. The HIM filling factor can
only be determined from observational data.

Four categories of hypotheses can be proposed to explain the
existence of expanding structures: 1) the sweeping of the ISM by
the winds of massive stars \citep{Loz92,Oey01}; 2) the progression
of an ionization front followed by a recombination \citep{McC87};
3) one or several SN explosions \citep{Ten87}; and 4) the
collision of high velocity clouds of neutral gas with the disk of
the Galaxy. The choice of the right hypotheses is difficult but
determinant, since the astrophysical implications regarding the
different ISM models derived from the existence of expanding
objects are completely different in each of the four above cases.
There exists no complete and general theory explaining all the
aspects of the stellar winds and SN interactions with the ISM.
Nevertheless, there are some simplified models that are able to
explain the observations \citep{Elm77,McC87,Loz92}.

What makes the choice of a model difficult is, firstly, that they
are all incomplete, and secondly, that the observational data are
too limited to supply enough typical cases to make a significant
validation. This last point in itself clearly justifies our
purpose: even if our automatic bubble detector could only find a
few very regularly shaped objects, this would be an appreciable
contribution to the discrimination of existing models of bubble
formation and evolution. The high resolution of the CGPS makes it
possible to visualize bubbles of radius down to a few parsecs in
the Perseus Arm. Those are especially ill represented in the
current database, and are thus not well studied even if they are
less complex than large scale structures. Considering the extent
of the CGPS field (666$^\circ$$^{2}$), the need of an automatic
detector for expanding shells is obvious.

\subsection{Proposed Strategy}

Little work has been published on the automatic detection of
expanding shells \citep{Mas99,Thi98,Mastbp}, and their aim was
mostly on the comparison of spherical structures observed in the
velocity channels of HI data cube, with structures predicted by
hydrodynamical simulations of expanding bubbles in the ISM. They
are therefore limited to morphological recognition.

Our approach is to recognize the \emph{dynamical behavior} of an
expanding bubble, characterized by its velocity spectra. The 21 cm
data collected by the CGPS contain HI abundance \emph{and}
kinematical information: as mentioned earlier, a characteristic
dynamical profile can in principle be found in every velocity
spectra extracted from every pixel of the bubble. This detection
technique avoids the difficulties linked to the often distorted
morphology of the bubbles. Bubbles are indeed hardly ever
spherical and most often incomplete, which implies the
subjectivity of a human observer and the difficulty of a purely
morphological detection. Working with the velocity spectra allows
a pixel-by-pixel evaluation of the likelihood of finding a bubble
at a particular location in the data cube. Subsequent steps are to
look for high concentrations of selected pixels in the data cube,
and finally to consider the morphology of the detection, which is
no longer the principal visual characteristic of an expanding HI
shell.

The presence of two peaks separated by 20 km s$^{-1}$ in a
velocity spectrum is unfortunately not  sufficiently restrictive
to make a reliable bubble detection. As will be discussed in
sections 3 and 4, it is partly for this reason that we used an
artificial neural network (ANN) for the first detection step. This
tool is now often encountered in astrophysics, mostly as a
classification device. We therefore present an overview of our ANN
classifier in section 3, after the description of the CGPS data in
section 2. In section 4 we expose the pixel-by-pixel detection
method for the velocity spectra, while section 5 deals with the
morphological validation of hypothesized bubbles. In section 6,
the detector is performed on a simulated purely turbulent HI gas
data cube. We present some preliminary results in section 7 before
concluding in section 8.

\section{The Data}

Many 21 cm surveys of the Galaxy were accomplished in order to
perform large scale study of the ISM: \citet{Hei74},
\citet{Wea73}, \citet{Col80} and in particular the CGPS
\citep{Nor96,Tay99,Hig99}. The CGPS aimed at systematically
observing the galactic plane, in the radio continuum and in the 21
cm HI line, between longitudes 75$^\circ$ to 145$^\circ$ and
latitudes -3$^\circ$ to 5$^\circ$. This project is the result of
contributions from Canadian university researchers and is carried
out at the Dominion Radio Astrophysical Observatory (DRAO).

The synthesis telescope is a 7-antenna interferometer oriented
east-west and with a maximum length of 604 m. This allows a
$\sim$1.0' resolution at 1420 MHz, while the previous large-scale
HI surveys have resolution poorer than 4'. Each antenna is about 9
m in diameter, which gives a synthesized field of view of
2.5$^\circ$ at 21 cm. Because of the 13 m minimum length of the
interferometer, the minimum sampling resolution is $\sim$
1$^\circ$ at 21 cm. A  26 m antenna was used to observe the larger
scales up to the survey dimensions. The total field covers
666$^\circ$$^{2}$ in 190 overlapped individual fields composing 36
5.12$^\circ$ mosaics. Due to the primary beam of each antenna, the
attenuation of the signal increases with the distance to the field
center. This and the partial overlap of the individual fields
causes the noise to vary in the final images. For this reason, a
noise level map is supplied for each mosaic. Table 1 summarizes
the parameters of the DRAO 21 cm HI data.

\section{Artificial Neural Networks (ANNs)}

An ANN is a network of many simple parallel operating elements
(see Figure 2). This configuration is inspired by the way
biological neurons effectively communicate: neurons located in
successive layers are massively interconnected by synaptic
connections allowing electric signal transmissions. One neuron
thus collects signals from many neighbors, and every signal is
weighted by the efficiency of the synaptic connection that is
involved. Learning can be defined technically as the process of
reinforcement and inhibition of those connections, in order to
create a neural network appropriate to the task to be learned. In
the case of ANNs, a free parameter called the \emph{synaptic
weight} stands for the connection efficiency.

A schematic representation of an example of a feedforward neural
network is showed in Figure 2. This network has a
three-dimensional input space, one hidden layer containing four
neurons, and its output layer contains one neuron - the output
space is one-dimensional. The neurons are represented as summation
signs because they compute the weighted sum of their synaptic
connections. To each connection is associated a synaptic weight,
which is an adjustable free parameter. The dimensions of the input
and output spaces, as well as the hidden layers dimensions, must
be determined by the user.

\subsection{The Backpropagation Algorithm}

ANN learning consists in the adjustment of all the synaptic
weights of all the connections, until a satisfactory mapping is
achieved. The so-called supervised learning of an ANN consists in
feeding the network with a training set of input-output mappings
$(\vec{x}(n),\vec{d}(n))$, where $\vec{d}(n)$ is the desired
output for the $n$th input vector $\vec{x}(n)$. After each
presentation, the ANN error is computed and is used to adjust the
free parameters, so that the error will be minimized if the same
situation is encountered. The learning process consists of
adjusting all the weights, so that  at each iteration the network
is heading for the direction of the more negative gradient on the
error surface. The usual learning algorithm used to achieve this
iterative minimization is called the backpropagation algorithm,
because the error at the network output is propagated backward
through the layers in order to calculate the proper adjustment for
each weight in each layer.

The backpropagation algorithm updates the weights for each neuron,
beginning with the output layer. Let $e_{j}(n)$ be the observed
error at the output neuron $j$ for the training datum $n$:

\label{bozomath}
\begin{equation}
e_{j}(n)=d_{j}(n)-y_{j}(n)
\end{equation}
where $d_{j}(n)$ is the desired output for neuron $j$ and
$y_{j}(n)$ ) is the observed output. The goal is to minimize
$E(n)$, the sum of the mean square errors (MSE) observed on the
set of $C$ output neurons for the training datum $n$:

\begin{equation}
E(n)=\frac{1}{2}\sum_{j \epsilon C} e_{j}^{2}(n)
\end{equation}
The output of neuron $j$ is defined by:

\begin{equation}
y_{j}(n)=\varphi[v_{j}(n)]=\varphi[\sum_{i=1}^{r}
w_{ji}(n)y_{i}(n)]
\end{equation}
where $\varphi[.]$ is the neuron transfer function (a usually
nonlinear function in order to produce a nonlinear mapping),
$v_{j}(n)$ is the weighted sum of the neuron $j$ inputs,
$w_{ji}(n)$ is the connection weight between the neuron $i$ in the
preceding layer (containing $r$ neurons) and the neuron $j$ in the
output layer, and $y_{i}(n)$ is the neuron $i$ output - see Figure
3.

To minimize $E(n)$, the weight $w_{ji}(n)$ must be updated in the
direction in which the error gradient
$\frac{\partial{E(n)}}{\partial{w_{ji}(n)}}$ is decreasing:

\begin{equation}
\triangle{w_{ji}}(n)=-\eta{\frac{\partial{E(n)}}{\partial{w_{ji}(n)}}}
\end{equation}
where $0 \leq \eta \leq 1$ is the learning rate. The chain rule
gives:

\begin{equation}
\triangle{w_{ji}}(n)=-\eta{\frac{\partial{E(n)}}{\partial{w_{ji}(n)}}}=\eta\delta_{j}(n)y_{i}(n)
\end{equation}

\begin{equation}
\delta_{j}(n)=e_{j}(n)\phi^{\prime}_{j}
\end{equation}
where $\delta_{j}(n)$ is called the local gradient of the error
surface.

The case of the hidden layers is similar, but the local gradient
of the neuron $j$ is now function of all the local gradients of
the neurons $k$ in the layer following the one who's neuron $j$ is
being updated:

\begin{equation}
\delta_{j}(n)=\phi^{\prime}_{j}\sum_{k \epsilon C}
\delta_{k}(n)w_{kj}(n)
\end{equation}

\subsection{Classification with ANN}

An ANN may be described as an algorithm that maps data, from an
input space into a user-chosen output space. Among other
possibilities, mapping a data set into an output space having
meaningful categories is equivalent to classification. Figure 4
illustrates the classification of data by mapping the input set
into a higher dimensional intermediate space, permitting the
separation of two categories. The data are afterwards mapped into
the output space as separated sets, or classes. It is this
particular application of the ANNs that we want to take advantage
of for this pattern recognition task in the velocity spectra.

One interesting thing is the generalization capability of the ANN,
which can interpolate between new and unknown input data. This
generalization ability, as well as the speed of the convergence
and the final ANN global performance, crucially depends on the
training set. For all the eventual input data to be properly
represented in the output space, the training set must cover most
of the space. It must include as various examples as possible, and
cover equally the different cases that may be encountered while
using the trained ANN. A very common case in the training set will
be very well delimited in the output space, while a rare one could
be associated to some other class or be bluntly treated as noise.

The main distinction between ANNs and more classical classifiers,
such as cross correlation, is thus the determination of the
classification criteria. A classical classifier needs precise
criteria that are sufficient to make a distinction between
classes.  Those criteria are initially known and chosen by the
user, hence they must have a known signification, or at least be
empirically identified and accessible.  To use an ANN
classification, one only requires a good and representative
training set. The ANN will, in a way, choose its classification
criteria by itself, by adjusting its synaptic weights. The only
known conditions are the ANN configuration (number of layers,
number of neurons, transfer functions) and the initial values of
the free parameters. The only a priori ``criterion'' is the error
minimization algorithm. What is obtained is therefore a totally
empirical classification, with no a priori statistical model.

This last point may cause some uneasiness: the fact that the ANN
chooses the distinction criteria may be seen as a loss of control
on the part of the user. An ANN could choose criteria that are
entirely irrelevant to the user's purpose, and consequently it
would be unable to process other data than the ones of the
training set. On the other hand, when intuitive criteria are not
available or are not sufficiently restrictive to achieve a good
classification, an ANN may be able to isolate hidden
characteristics and use them to make relevant decisions. Moreover,
its generalization ability allows the recognition of distorted and
noisy data, that a classical classifier might not be able to
process.

These last two points justify our use of an ANN for the detection
of the kinematical signature of expanding HI bubbles. Another
classifier could have been used, but based on uncertain
classification criteria: for example, cross-correlation would
require a spectral template that we do not possess. Indeed, the
search of a profile with peaks separated by 20 km s$^{-1}$ is not
sufficiently restrictive.  Our detector is thus required to find
other criteria to recognize the bubbles' velocity spectra. The
data are furthermore noisy - in addition to the detection noise,
we have to deal with structural noise: the expanding shells evolve
in a HI environment made of clouds of various sizes, of filaments,
and other various structures that disturb the spectra. We believe
that an ANN classifier may be more likely than any other
classifier to integrate a sufficiently general and restrictive
model of an expanding bubble velocity profile, in order to
recognize it amongst all other structures, and despite the noise.

There are several possible ANN configurations. We chose a
multilayer perceptron (MLP), which is a feedforward neural network
such as the one described by Figure 2, because it proved its value
in many practical classification problems, and also because it
does not make any a priori assumption about data distribution,
contrary to other statistical classifiers. The main shortcoming of
the MLP in this particular context is that it may take decisions
about regions of space that were not covered by the training set.
The MLP will not reject a case pleading incompetence. Independent
rejection criteria allowing to discard some of the MLP decisions
will be described in sections 4 and 5.

The main difficulty of our proposal is about the training set. The
small number of known certified observational bubbles makes it
difficult to assemble a set of diversified and representative
examples. This certainly requires special attention, as will be
discussed later.

For more details about the theory and implantation of ANNs, the
reader can consult: \citet{Hay98}, \citet{Dem00}, \citet{Pri00}.

\section{Dynamical Detection}

The detection of the dynamical characteristics of a bubble in the
velocity spectra is a pixel-by-pixel one: the goal is to assign a
degree of confidence to each pixel, expressing how much the
corresponding velocity spectrum matches the dynamical profile that
is looked for. This first step is taken care of by a MLP with two
hidden layers (the layers between the input and output layers)
containing 17 and 2 neurons, respectively. The error minimization
is performed by a Levenberg-Marquardt backpropagation algorithm,
which is a faster variation of the standard backpropagation
algorithm exposed in section 3 \citep{Hag94}. The learning rate is
0.17. Those conditions allow to reach a good MLP performance (mean
square error (MSE) $\leq$ 0.014) in a relatively short period of
time. The MATLAB Neural Network Toolbox has been used.

Too long a training can compromise the MLP generalization
capability: the worst limit is the MLP having learned to recognize
each individual training spectrum, and being unable to recognize
any other similar spectrum that is unknown. A standard
cross-validation technique that averts overtraining is the use of
a validation set, which is a data set composed as the training
set, but that is not used for the training itself. The MLP
performance is computed for each iteration over the validation
set, and the training is stopped when the error over the
validation set begins to rise. The training set contains
$\sim$1300 spectra and the validation set $\sim$200. This ratio is
standard.

The vectors presented at the input layer of the MLP are 12-channel
segments of the velocity spectra. This number of channels spans
$\sim$20 km s$^{-1}$, a velocity interval sufficient to contain
the dynamical profile of the bubble. The background (an average
spectrum over the whole data cube) is subtracted beforehand, in
order to filter the irrelevant global characteristics such as the
local diffuse material associated with the spiral arm. Only the
local HI enhancements and dynamical characteristics (such as an
expanding bubble) are preserved. The MLP output is a number in the
range [0, 1] expressing how good the 12-channel segment matches
the dynamical profile of an expanding bubble. Figure 5 shows
examples of spectra classified by the MLP. Even if the three
examples show a two-peaked profile, the MLP output is different
for each. Some information that is not directly accessible or
intuitive, but that allows some discrimination, is therefore
extracted by the MLP.

The results presented in this paper were obtained with a committee
of 4 MLPs. We simply averaged the results coming from each of the
4 networks. The MLPs were trained over velocity spectra taken from
6 different data cubes, each of which containing one known
bubble.Two of those bubbles are artificial bubbles: spheres of
relatively high brightness temperatures have been added in real
and otherwise unaltered HI data cubes. This is a way to enrich and
improve the training set by making it more representative: even
though those two bubbles are not real, inserting them into the
training set introduces representative dynamical characteristics
inside various HI environments, and thus contributes to enrich
knowledge of the MLP. The 4 real bubbles are: G132.6-0.7-25.3; the
bubble related to WR144 (galactic coordinates 80.04+0.93-26.2);
the bubble related to WR139 (76.6+1.43-69.07); and the bubble
related to WR149 (89.53+0.65-66). The MLP performance will
naturally get better as the training set will be enriched with new
bubble instances.

As mentioned in section 3.2, a MLP is not well suited for
rejecting data: it will output a decision for every input datum,
even if it is unable to make a proper classification given the
experience acquired during its training. The MLP taken alone may
thus make an important number of false detections: generally, 2\%
to 10\% of the pixels are incorrectly identified ``bubble'' in the
whole data cube. A filtering is therefore performed by verifying a
number of conditions in each selected spectrum. By comparing our
known examples of positive spectra (spectra belonging to known
bubbles), we have been able to isolate some discriminant features
most of them have in common. However, the negative spectra may
have every possible shapes, and occasionally present some of these
features. For this reason, we rather identify \emph{rejection
criteria}, allowing us to exclude a detection made by the MLP.
These conditions are supplied by statistics compiled over all the
positive cases we have. Those characteristics are very pragmatic,
such as the values and positions of the spectrum extrema. Applying
those rejection criteria on the MLP detections allows us to
exclude a significant part of the false detections, but keeps most
of the correct ones.

Figure 5 a) illustrates a typical positive spectrum which shows
some positive visual characteristics:\\ $\bullet$ The minimum position is between channels 6 and 9; \\
$\bullet$ The minimum value is $<$ -12 (negative values are due to
the
background subtraction); \\
$\bullet$ The maximum value is $>$ 0. \\ The approximately central
position of the minimum is a quite intuitive feature that depends
directly on the dynamics of expanding bubbles. However, the values
of the extrema (which are brightness temperatures relative to the
average background) depend on physical conditions that are less
stable, such as the ratio between the shell and ambient medium
densities. In fact, the rejection conditions were chosen on purely
empirical basis: we only noticed they could make some distinction
between a positive set and a negative set. Figure 6 shows the
statistics of the 3 characteristics stated above, to visually
compare positive and negative sets. One can notice some trend for
the positive sets. If some permitted intervals are established for
each condition, an important part of the false detections can be
rejected. In the examples presented in section 7, we simply
established lower and upper limits such that 88\% of the spectra
of the positive histograms are preserved.

\section{Morphological Detection}

We believe the pixel-by-pixel dynamical detection is helpful to
begin the search because of the difficulties arising from a
morphological detection in several successive velocity channels.
This first step taken, we are left with a cube in which patches of
positively identified pixels are found. In a single velocity
channel, this results in connected objects, or ``blobs'', of
various dimensions and shapes.

\subsection{The Dimensions of the Object}

The object dimension in pixels depends on its intrinsic dimension
and on its distance. As a velocity channel is associated to a
galactocentric distance, it is possible to establish a lower
limit, function of the velocity channel, to the number of pixels
corresponding to a blob of a few parsecs diameter, and thus
necessary for a blob to be a real bubble. The higher limit for the
bubbles' size is indirectly determined by the dynamically based
detection, since the maximum expansion velocity for bubbles of a
few parsecs is typically 10 km s$^{-1}$.

\subsection{The Object Shape}

One should keep in mind that the blobs with which we now have to
work are in fact \emph{concentrations of positive detections} -
that is to say, their shape do not necessarily match the visual
shape of the bubble. Despite this fact, it is clear that a
concentration of detections having a roughly circular and
condensed shape is more likely to be a bubble, than some very
elongated and sharp filament. It is thus relevant, at this point,
to take morphology into account.

To begin with, the size and orientation of the object are
characterized using a principal component analysis: the directions
of the two principal axis of the blob are calculated, and the
eccentricity of an associated ellipse is estimated by the ratio of
the eigenvalues of the axis (see Figure 7). A new rejection
criteria is then defined based on a superior limit to the
eccentricity.

\subsection{Increasing Intensity Criterion}

Because an HI bubble corresponds to a local lack of gas, another
rejection criterion for a given velocity channel, directly
linkable to the physics of the phenomenon, and independent of the
dynamics or morphology, is the fact that the intensity of the 21
cm emission should increase in a radial fashion around its center
(see Figure 8).

\section{Turbulence: Simulation of self-similar HI in the Perseus arm}

As mentioned earlier, the structural noise due to various
structures in HI may cause false detections. Apart from the
rejection criteria and the morphological confirmation just
exposed, there is not much to be done against it. It is however of
interest to verify if our detector could be tricked by turbulence
alone. We performed our detector on a simulated purely turbulent
HI gas data cube such as described below. The subsequent rejection
criteria on the velocity spectra were not used, because the
brightness temperatures relative to the average background could
not be consistent in real and simulated cubes. The committee of
MLPs does retain some spectra as positive, but those detections do
not pass the morphological test: in fact, no blob at all can be
isolated, the maximum number of connected pixels in a
concentration of positive detections being 18 ($\sim$$\frac{1}{4}$
of minimum number accepted by the detector).

The rest of this section describes how we have simulated the
spectro-imagery observation of the observed part of the Perseus
arm on the hypothesis that the HI density structure and kinematics
are only due to turbulence and Galactic rotation. In this section
we describe how we modelled the HI 3D density and velocity fields
and how we constructed the spectro-imagery observation from these.

We consider that the center of the Perseus arm is at a distance of
2 kpc from the Sun and that its depth is 0.3 kpc. As the field
observed at DRAO is $2.6^\circ\times2.6^\circ$, this translates
into a surface of $\sim0.1\times 0.1$ kpc at the Perseus arm
distance. To simulate the Perseus arm we have produced a
$128\times128\times 384$ box that represents the observed portion
of the Perseus arm ($0.1\times0.1\times0.3$ kpc). Each cell in
that box has a linear size of 0.75 pc.

\subsection{The 3D velocity field}

The simulated 3D velocity field ($v(x,y,z)$) is the sum of the
Galactic rotation component ($V(x,y,z)$) and of a turbulent
component ($\delta v(x,y,z)$). To lighten the text, we drop the
$(x,y,z)$ in the following.

\subsubsection{The Galactic rotation velocity component}

To simulate the Galactic rotation velocity component, we proceeded
as follows. Each point in the simulated cube has been attributed a
longitude $l$, latitude $b$ (corresponding to our DRAO
observation) and a distance from the sun (the center of the box
being at 2 kpc). The Galactic radius $R$ of each cell element is
then given by:
\begin{equation}
R = ( D\cos(l)-R_\odot ) / \cos(\theta)
\end{equation}
where
\begin{equation}
\theta = \tan^{-1}\left(\frac{D\sin(l)}{D \cos(l)-R_\odot}\right)
\end{equation}
with $R_\odot=8.5$ kpc.

Following  \citet{burton92}, the Galactic angular velocity
$\Omega$ has been estimated to be:
\begin{equation}
\Omega = \Omega_\odot  (1.0074*(R/R_\odot)^{0.0382} + 0.00698 )
\end{equation}
with $\Omega_\odot = 220$ km s$^{-1}$. Finally the angular
velocity component along the line of sight is given by the
following expression:
\begin{equation}
V = (\frac{R_\odot}{R} \Omega - \Omega_\odot) \times \sin(l)
\cos(b).
\end{equation}

\subsubsection{The turbulent velocity component}

To simulate the self-similar properties of the velocity structure
we have used 3D fractional Brownian motion simulations
\citep{stutzki98,miville-deschenes2003}. We have simulated a
$128\times128 \times 384$ cube with spectral index -11/3, to
simulate Kolmogorov type turbulence. The average of the turbulent
velocity field was set to zero ($<\delta v>_{x,y,z}=0$) and its
dispersion to 20 km s$^{-1}$.

\subsection{The density field}

The simulated 3D density field ($n(x,y,z)$) was also built using a
fractional Brownian motion simulation with a spectral index of
-11/3. No correlation between velocity and density was introduced.
We have subtracted the minimum value from the 3D density cube to
allow only positive density values. Because of the relatively
small size of the observed region, we did not introduce any
variation of the average density with the height above the
Galactic plane.

\subsection{Building the spectro-imagery observation}

The spectro-imagery observation (also called
position-position-velocity or PPV cube) has been constructed by
making the assumption that every gas cell on the line of sight
emits a thermally broaden Gaussian ($\sigma=\sqrt{ k_B T/m }$)
centered at $v(x,y,z)=V(x,y,z)+\delta v(x,y,z)$ and of amplitude
$n(x,y,z)$ (see \citet{miville-deschenes2003} for the details). We
have considered optically thin Cold Neutral Medium gas at $T=100$
K. We have constructed a $128\times128\times128$ PPV cube that
gives the column density on each line of sight and at each
velocity $u$ (velocity bin is 1.65 km s$^{-1}$) using the
following equation:
\begin{equation}
\label{eq_PPV} N_H(x,y,u)=\alpha \sum_z n(x,y,z) \exp \left (
-\frac{[u-v(x,y,z)]^2}{2\sigma^2} \right ).
\end{equation}
Here $\alpha$ is a normalization factor that takes into account
the spectral resolution and the linear size of a cell along the
line of sight.

\section{Preliminary Results}

The two test cubes have galactic coordinates ($\Delta$l;
$\Delta$b; $\Delta$v) = (136.75, +139.30; -0.10, -2.65; +12.6,
-136.7) and ($\Delta$l; $\Delta$b; $\Delta$v) = (73.1, 75.7;
+0.25, +2.9; +60.7, -102.9), and contain respectively bubbles
GSH138-01-094 and G73.4+1.55. Those two bubbles were not used for
the MLP training, therefore the detector possesses no preliminary
knowledge of those particular cases. It is believed that
GSH138-01-094 could have been caused by a SN \citep{Sti01}, while
G73.4+1.55 is linked to WR134 \citep{Ger99}.

It took a few hours to train the MLPs, and their MSE on the
validation set is $\leq$ 0.014. The detection itself on a cube of
128x128x79 12-channel spectra takes $\sim$ 5 minutes with a
Pentium 4, 1.9 GHz and 512 Mo ram.

Figure 9 shows the velocity channel corresponding to v $\approx$
-94 km s$^{-1}$, where GSH138-01-094 should be detected, through
the stages of the detection. Figure 10 shows the same stages for
G73.4+1.55, at v $\approx$ -11.4 km s$^{-1}$. Table 2 traces the
evolution of the amount of false detections for each step, in
pixel percentages and in number of preserved blobs.

Figure 11 shows a detection at (l,b,v) $\sim$ (138, -0.5, -7) in
the same data cube where GSH138-01-094 is found. The presence of a
bubble is visually very convincing, as much in its morphology
through successive velocity channels, as in its dynamical
signature in the velocity spectra. It would certainly be relevant
to check the eventual presence of a WR star or a supernova remnant
at this location.

\subsection{Discussion}

0.5\% to 0.7\% of the detections, or less than 30 to 60 blobs, do
not correspond to any known bubbles. It is however tempting to
verify those false detections in the hope of finding out that some
of them were justified. What we want to point out is that, even if
our detector gives several false alarms, those very limited and
confined regions of the sky constitute a search field considerably
reduced compared to the whole sky. The purpose of our detector is
not to make a final decision about the validity of a detected
bubble, but to raise flags in order to guide the researchers
towards regions of the sky with relatively high probability of
finding one.

\section{Conclusion}

The next step to this work is to use the detector on a large
section of the CGPS HI data. The eventual new cases of bubbles
found using the detector will be integrated to the MLP training
set in order to improve its specialization. For the first time, an
evaluation of the amount and distribution of expanding bubbles of
a few parsecs radius will be possible.

Such an evaluation is critical to evaluate the relative importance
of the HIM in the ISM, which is traditionally addressed in terms
of the porosity parameter $Q$. This parameter is defined as the
ratio between the volume occupied by the bubbles in the Galaxy and
the total volume of the Galaxy. $Q$ is thus directly related to
the HIM filling factor, if we suppose that the HIM fills the
bubbles \citep{McK77}. For example, \citet{Oey97} deduced $Q$ from
an analytical expression of the distribution and size of bubbles
and superbubbles. If their prediction for a maximum in the HI
cavities radii distribution was verified, this would constitute a
new constraint for the SN remnants evolution, for the ISM
conditions and for the typical energy injected by SN in galaxies.
However, an observational verification of this prediction in the
small radii regime is not possible because of the lack of known
cases. Our results could help to clarify this issue.

In this paper, it has been shown that a detector using the
dynamical features in the velocity spectra of HI data cubes can
point out zones where there is a relatively high probability of
finding an expanding bubble. This detector is based on the
classification of a MLP, and on some independent dynamical and
morphological rejection criteria. The test bubbles have been
correctly detected with a false detections percentage of $\sim
0.6\%$, which significantly reduces the subsequent search field.
This preliminary cleaning could be followed by a human inspection
or other more complex automatic analysis processes, which could be
applied on a few objects rather than a vast amount of rough data.

\acknowledgments

We thank Nicole St-Louis who provided us with the 21 cm data of
G73.4+1.55, obtained at DRAO, and Sergei Mashchenko for his
artificial bubbles code. This research was funded by the Natural
Sciences and Engineering Research Council of Canada, the Fonds
FCAR of the Government of Qu\'{e}bec, and Universit\'{e} Laval.

\clearpage

\begin{figure}
\plotone{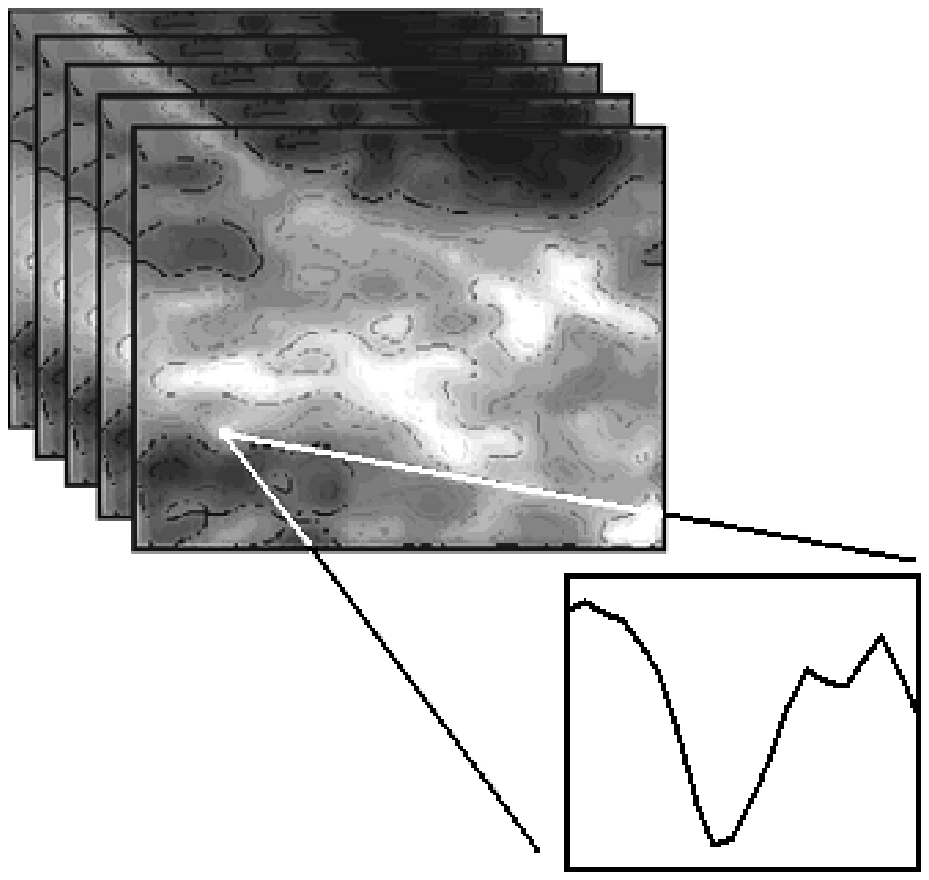}
\caption{Velocity spectra extracted from a 21 cm data cube.
Each velocity channel is an image of the HI gas at a given velocity.}
\end{figure}

\begin{figure}
\plotone{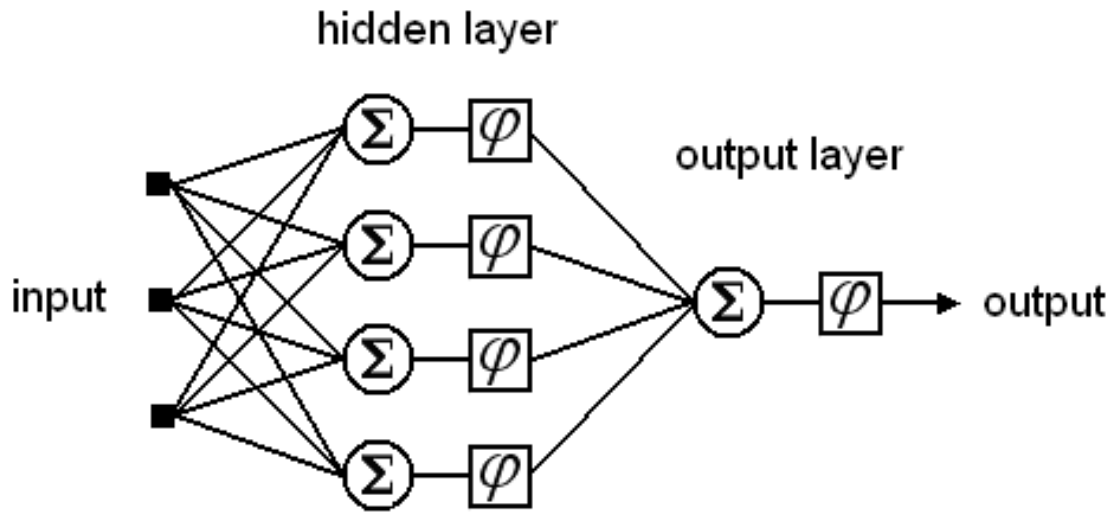}
\caption{Schematic representation of a feedforward neural
network. To each connection is associated a synaptic weight (the
free parameters). A neuron first computes the weighted sum of its
synaptic connections before applying a usually non-linear transfer
function $\varphi$ to produce an output. The dimensions of the
input and output spaces are at the user's discretion. In this
paper, the input vector (the input data) is a 12-component segment
of a velocity spectrum, and the output is a one-dimensional degree
of membership to the class of expanding bubbles.}
\end{figure}

\begin{figure}
\plotone{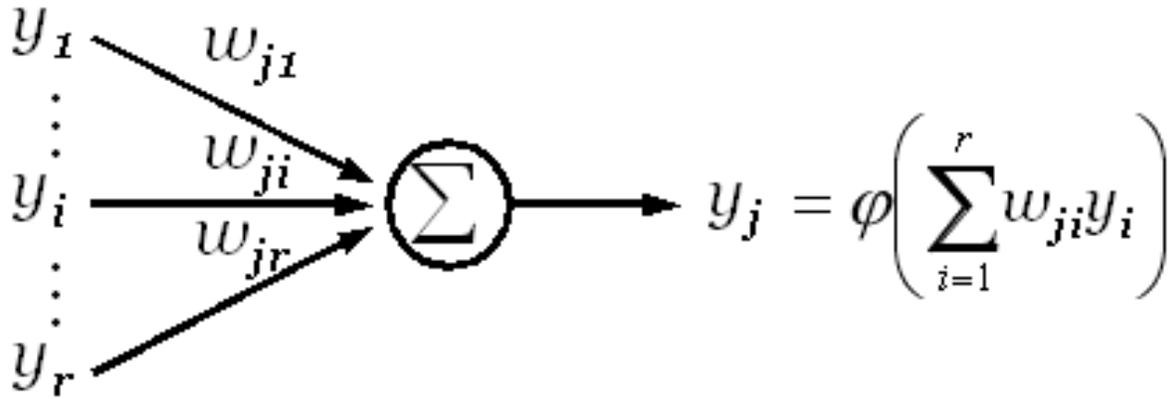}
\caption{The neuron $j$ in the output layer sums the weighted
outputs of the neurons $i = [1,r]$ of the preceding layer. The
transfer function $\varphi$ is then performed to produce the
neuron $j$ output $y_{j}$.}
\end{figure}

\begin{figure}
\plotone{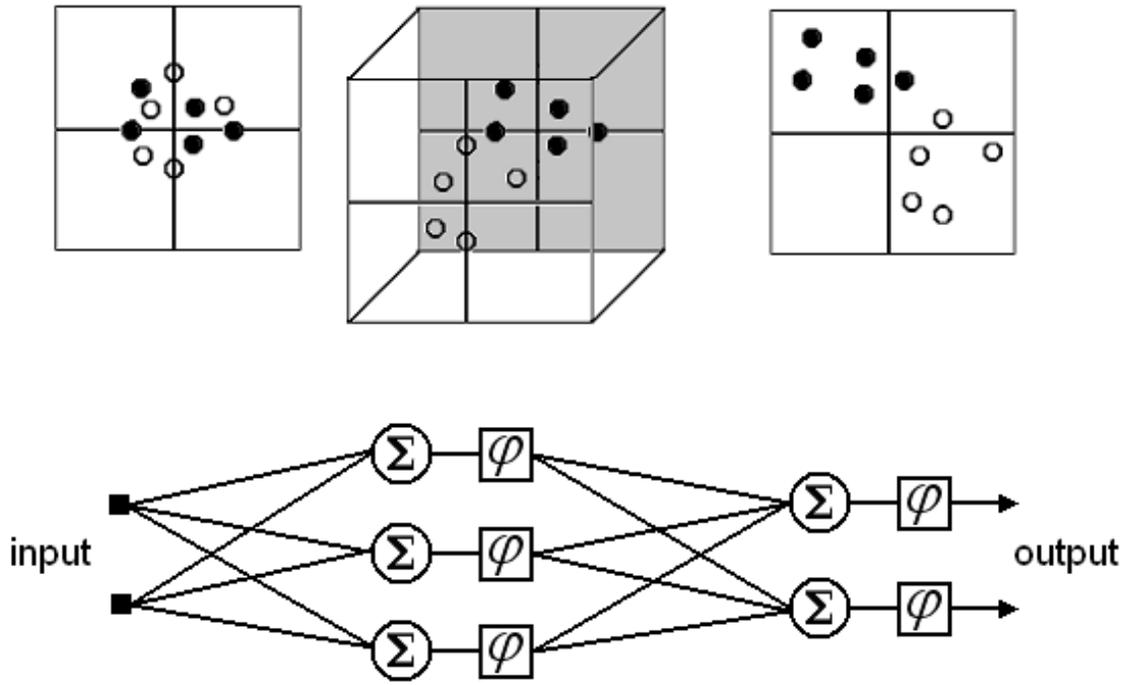}
\caption{Mapping of data from an input space into an output
space, in order to achieve a classification. The ANN 3-neuron
\emph{hidden layer} acts as a 3-dimensional intermediate space
allowing the separation of intricate data.}
\end{figure}

\begin{figure}
\plotone{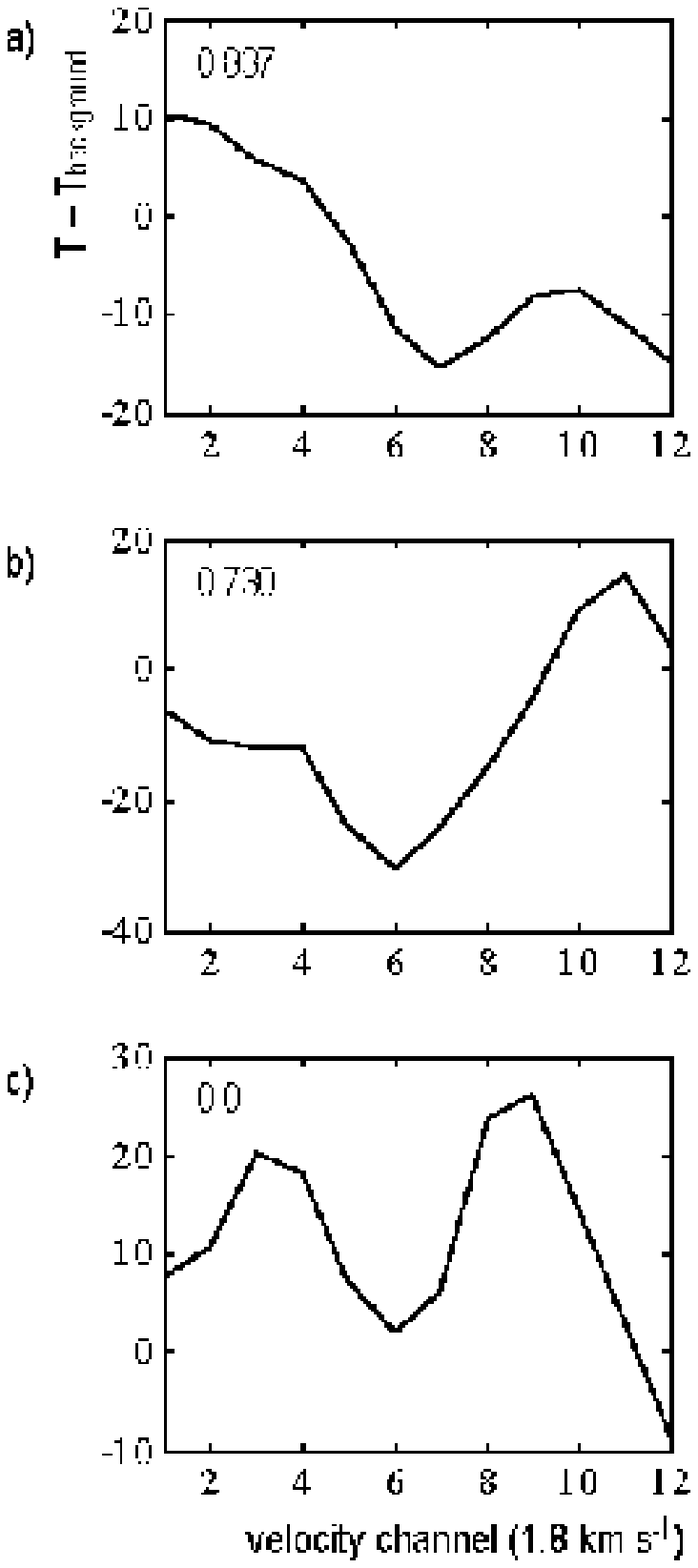}
\caption{Three velocity spectra. The MLP output is indicated on
each graph. a) Spectrum effectively positive and detected as such
(belongs to GSH138-01-094); b) Negative spectrum, insufficient MLP
output; c) Negative spectrum, insufficient MLP output.}
\end{figure}

\begin{figure}
\plotone{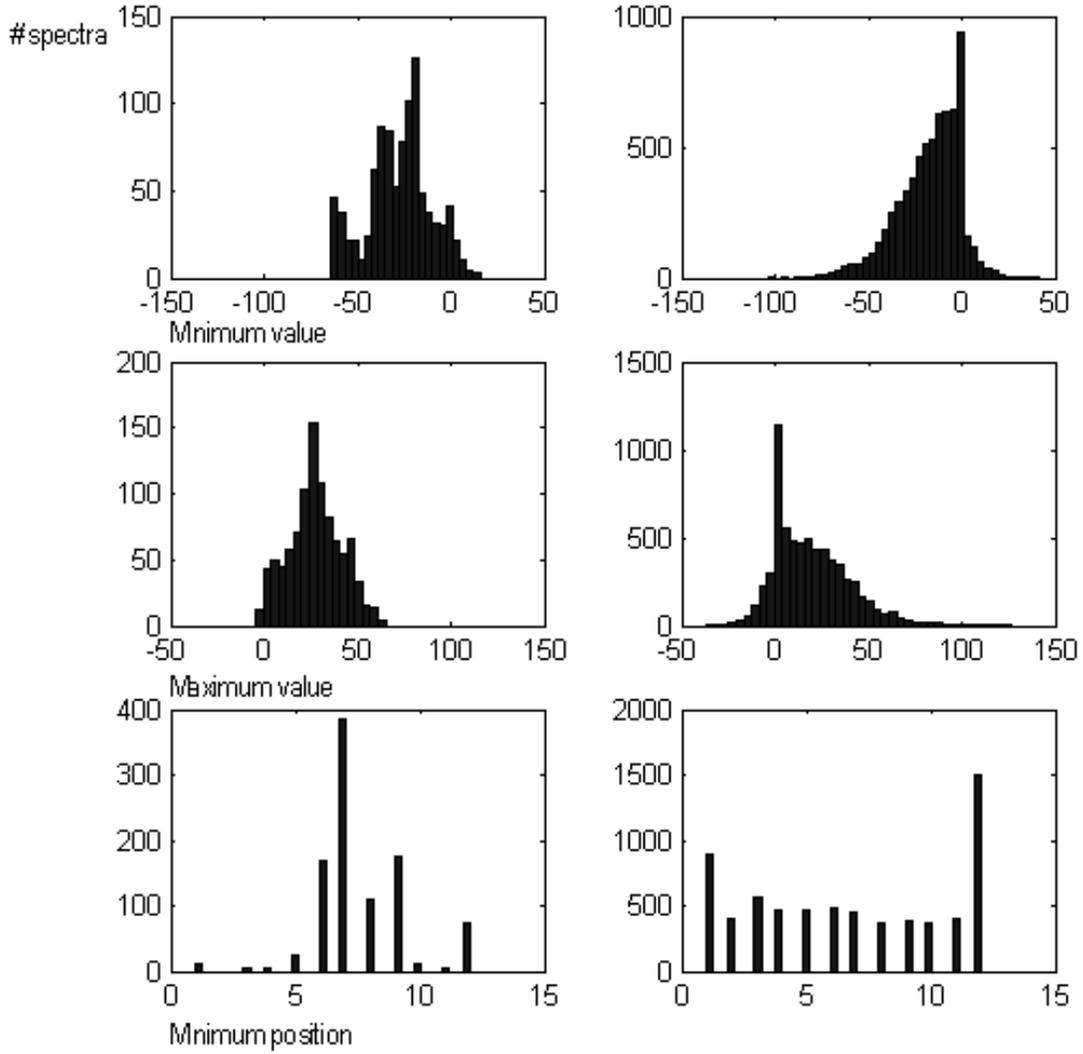}
\caption{Histograms illustrating the statistics of 3 visual
features in positive (left hand side) and negative (right hand
side) spectra sets. Downward : the value of the minimum; the value
of the maximum; the position of the minimum.}
\end{figure}

\begin{figure}
\plotone{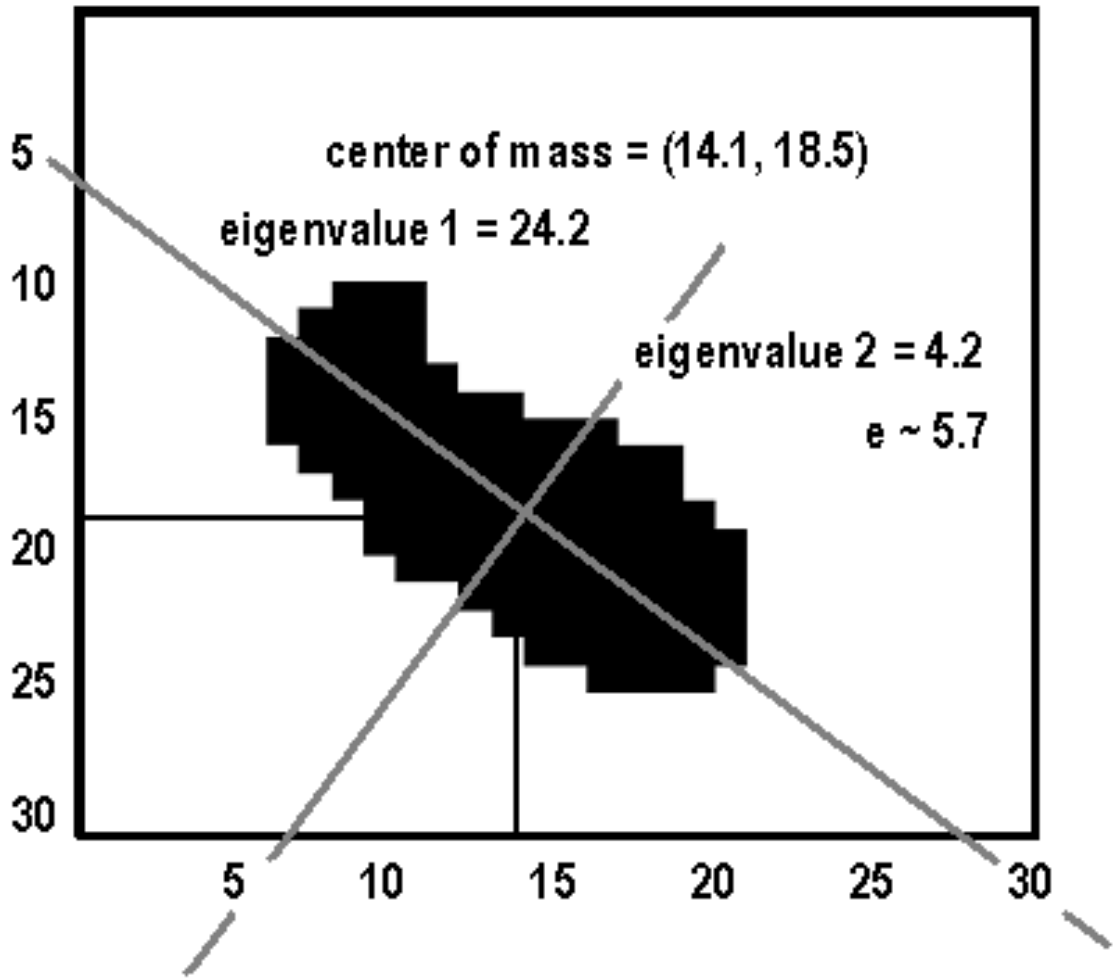}
\caption{Characterization of a blob by its association with an
ellipse. The ellipse eccentricity is estimated by the ratio of the
eigenvalues of its 2 principal axis.}
\end{figure}

\clearpage

\begin{figure}
\plotone{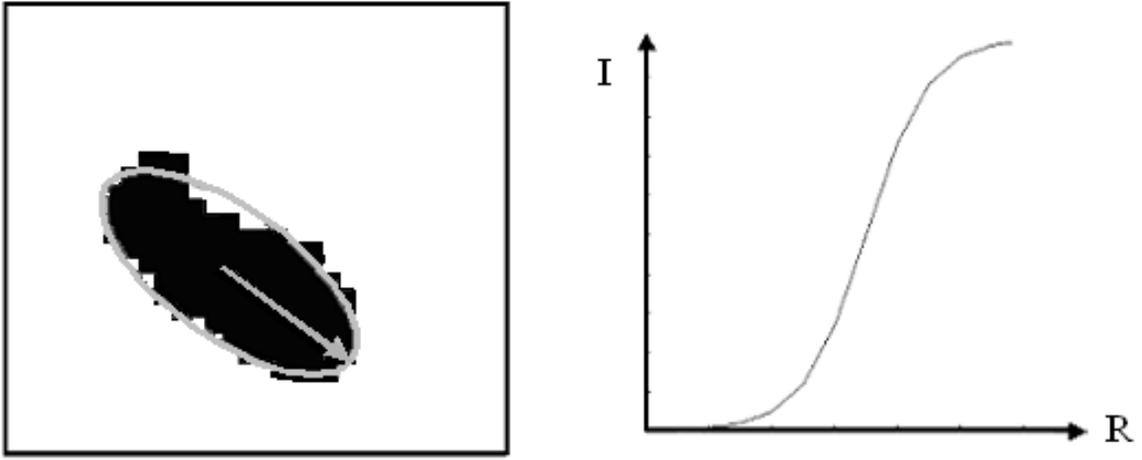}
\caption{The increasing intensity criterion : the HI abundance
must increase with the bubble radius. What is effectively verified
is that the pixels intensity on the ellipse principal axis
increases as we progress away from the blob center. This ``center"
is not the center of mass but the minimum of the intensity
neighboring the center of mass.}
\end{figure}

\begin{figure}
\plotone{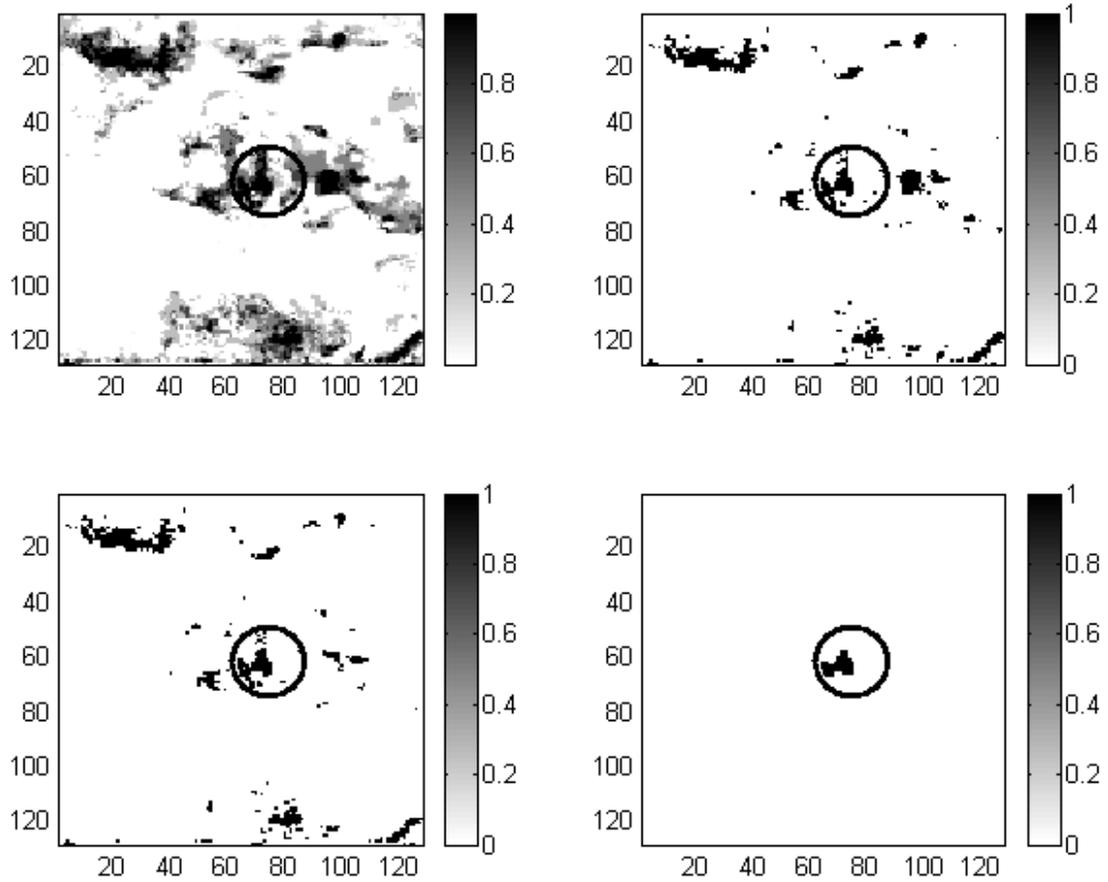}
\caption{Results of the detection of GSH138-01-094. The ellipse
indicates the bubble location and scale at its maximum expansion.
From left to right, downward : the MLP detection; after a 0.82
threshold; after application of the rejection criteria on the
spectra; after application of the rejection criteria on the
blobs.}
\end{figure}

\begin{figure}
\plotone{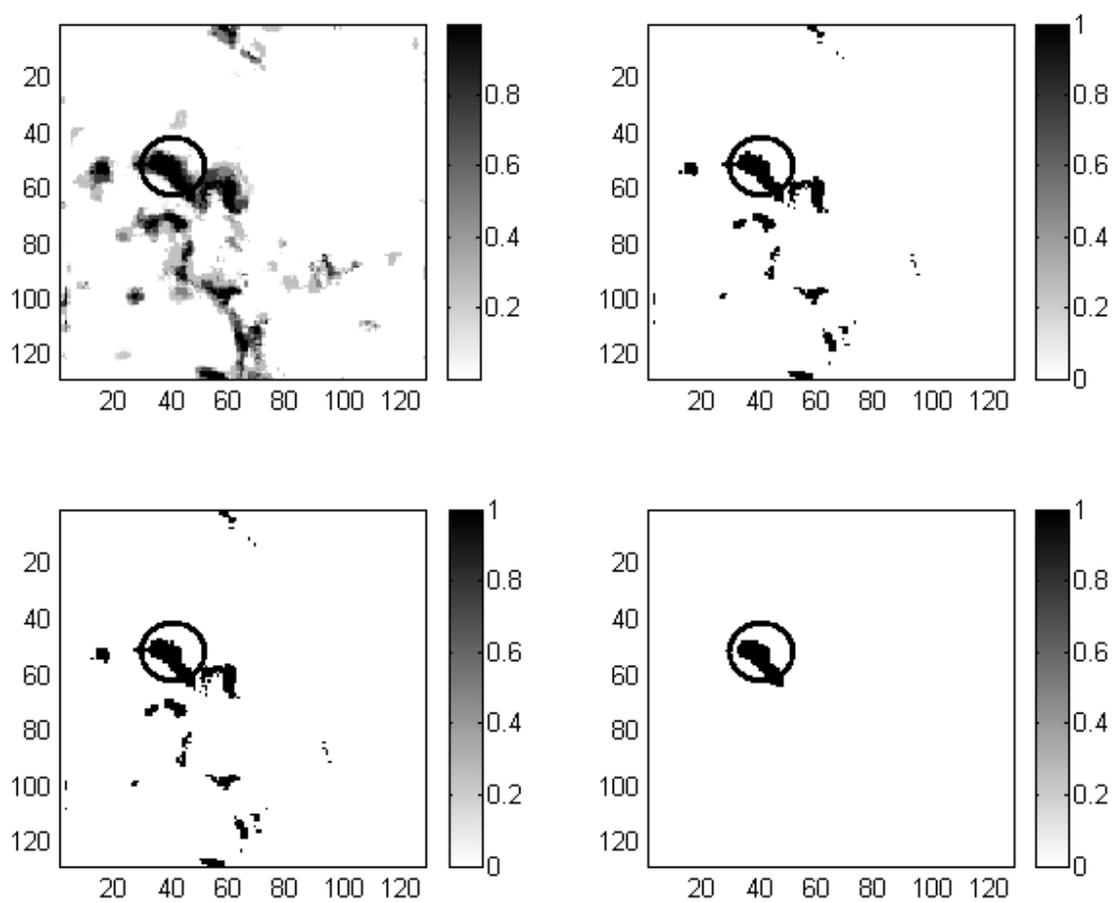}
\caption{Results of the detection of G73.4+1.55.}
\end{figure}

\begin{figure}
\plotone{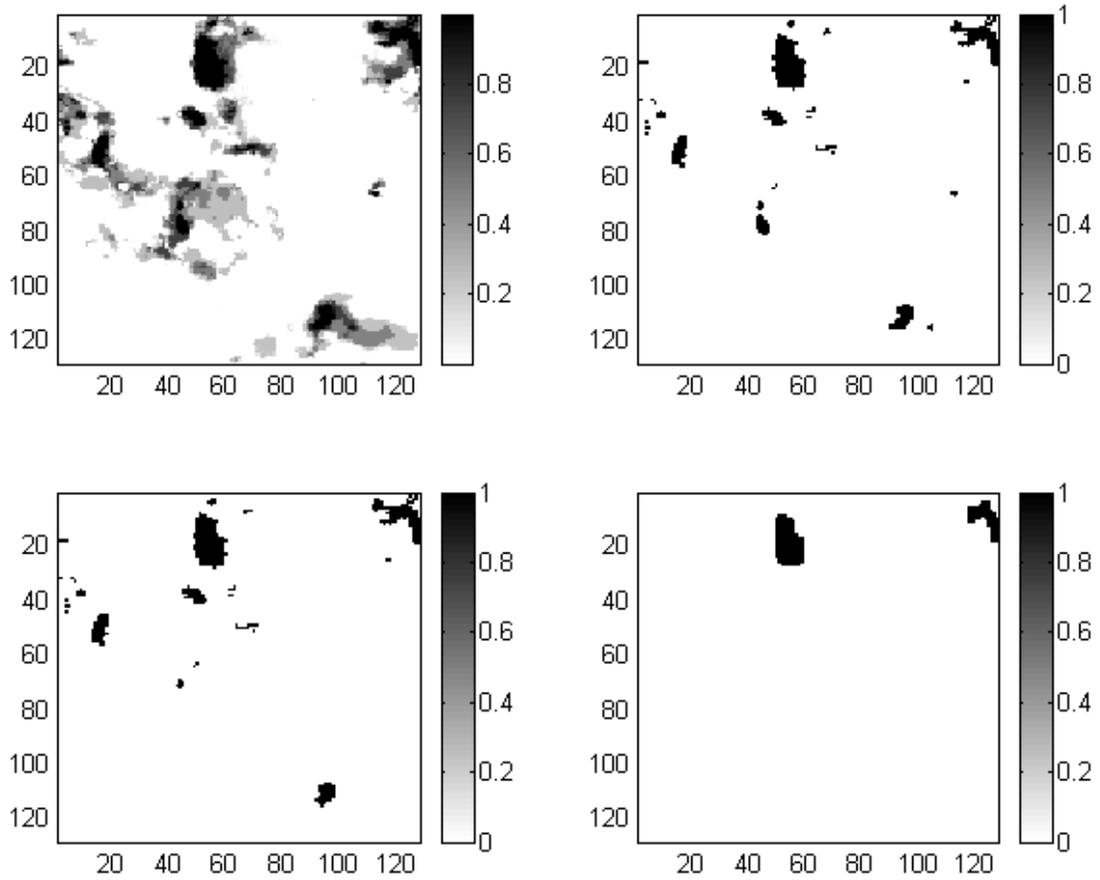}
\caption{Detection at coordinates (l,b,v) = (138, -0.5, -7).
Corresponds to the location of an unknown bubble, at least
visually convincing in its morphology as well as in its dynamics.}
\end{figure}

\clearpage

\begin{deluxetable}{cc}
\tabletypesize{\scriptsize} \tablecaption{Parameters of the CGPS
21 cm survey \citep{Hig99}. \label{tbl-1}} \tablewidth{0pt}
\tablehead{ \colhead{Parameter} & \colhead{Value} } \startdata
\textit{Survey} & \\
Number of Synthesis Telescope Fields &190 \\
Field Dimensions ($^\circ$) &2.5 \\
Spacing of Field Centers ($^\circ$)  &1.86 \\
Spatial Resolution &1.0'x1.0'$\csc\delta$ \\
Number of Velocity Channels &272 \\
Spectral Resolution (km s$^{-1}$) &1.319 \\
Channel Separation (km s$^{-1}$) &0.82446 \\
RMS Noise (Field Center) (K) &2.9 \\
\textit{Mosaics} &  \\
Number of Mosaics &36 \\
Size of Mosaics ($^\circ$) &5.12 x 5.12 \\
Overlap of Mosaics ($^\circ$) & 1.1 \\
Area Covered ($^\circ$$^{2}$) &666 \\
Galactic Longitude Coverage ($^\circ$) &74.2 to 147.3 \\
Galactic Latitude Coverage ($^\circ$) &-3.6 to 5.6\\
\enddata

\end{deluxetable}

\clearpage

\begin{deluxetable}{lcccc}
\tabletypesize{\scriptsize} \tablecaption{Evolution of the False
Detections Through the Detection Stages. \label{tbl-1}}
\tablewidth{0pt} \tablehead{ \colhead{Stages of the Detection} &
\colhead{GSH138-01-094} & \colhead{} & \colhead{G73.4+1.55} &
\colhead{}}
\startdata
   &\% False Detections(pixels)&\# Preserved Blobs&\% False Detections(pixels)&\# Preserved Blobs\\
Threshold of 0.8 &2.3 &375 &3.0 &475 \\
After Filtering the Spectra &1.9 &305 &2.7 &456 \\
After Filtering the Blobs &0.5 &29 &0.7 &57 \\
\enddata

\end{deluxetable}

\clearpage

\end{document}